\documentclass[amsmath,amssymb,11pt,superscriptaddress,reprint, preprintnumbers, notitlepage,aps,twocolumn,nofootinbib]{revtex4-1}
\pdfoutput=1 
\usepackage[utf8]{inputenc}
\usepackage[english]{babel}
\usepackage{amsmath}
\usepackage{graphicx}
\usepackage{dcolumn}
\usepackage{pbox}
\usepackage{amssymb}
\usepackage{epsfig}
\usepackage{slashed}
\usepackage{amssymb}
\usepackage{ mathrsfs }
\usepackage[usenames,dvipsnames]{xcolor}
\usepackage[font=small]{caption}
\usepackage[font=small]{subcaption}
\usepackage{url}

\usepackage{multirow}
\usepackage{lineno}

\renewcommand\fbox{\fcolorbox{black}{white}}

\definecolor{MyLightBlue}{rgb}{0.22,0.51,0.9}
\definecolor{BrickRed}{rgb}{0.8, 0.25, 0.33}
\RequirePackage{hyperref}
\hypersetup{colorlinks, citecolor=blue,linkcolor=BrickRed, urlcolor=MyLightBlue}

\makeatletter
\renewcommand\@makecaption[2]{%
  \par
  \vskip\abovecaptionskip
  \begingroup
  
   \small\rmfamily
    \begingroup
     \samepage
     \flushing
     \let\footnote\@footnotemark@gobble
     \@make@capt@title{#1}{#2}\par
    \endgroup
  \endgroup
  \vskip\belowcaptionskip
}
\makeatother

\DeclareUnicodeCharacter{2212}{-}
\begin{document}
\title{\Large 
Marriage between neutrino mass and flavor anomalies
}

\author{\bf J. Julio}
\email[E-mail: ]{julio@brin.go.id}
\affiliation{National Research and Innovation Agency, Kompleks Puspiptek Serpong, South Tangerang 15314, Indonesia}

\author{\bf Shaikh Saad}
\email[E-mail: ]{shaikh.saad@unibas.ch}
\affiliation{Department of Physics, University of Basel, Klingelbergstrasse\ 82, CH-4056 Basel, Switzerland}

\author{\bf Anil Thapa}
\email[E-mail: ]{wtd8kz@virginia.edu}
\affiliation{Department of Physics, University of Virginia, Charlottesville, Virginia 22904-4714, USA}

\begin{abstract}
Experimental hints for lepton flavor universality violation in beauty-quark decay both in neutral- and charged-current transitions require an extension of the Standard Model for which scalar leptoquarks (LQs) are the prime candidates. Besides, these same LQs can resolve the long-standing tension in the muon and the recently reported deviation in the electron $g-2$ anomalies. These tantalizing flavor anomalies have discrepancies in the range of $2.5\sigma-4.2\sigma$, indicating that the Standard Model of particle physics may finally be cracking. In this Letter, we propose a resolution to all these anomalies within a unified framework that sheds light on the origin of neutrino mass. In this model, the LQs that address flavor anomalies run through the loops and generate neutrino mass at the two-loop order while satisfying all constraints from collider searches, including those from flavor physics. 
\end{abstract}

\maketitle
\section{Introduction}
In the Standard Model (SM), the coupling of the electroweak (EW) gauge bosons to leptons is flavor universal, a property known as lepton flavor universality (LFU). The SM's extension is expected to violate this property, and experimental observations of such processes will be direct evidence of physics beyond the SM (BSM). Even though LHC searches have not found any direct hints for new particles, indirect searches for new physics have become an increasingly important pathway. Intriguing hints for BSM physics have accumulated over the years from the measurements of precision observables in several experiments that strongly suggest LFU violation (LFUV).

Lepton anomalous magnetic moments (AMMs) for the first two generations that are very sensitive to new physics are measured in the experiments with unprecedented accuracy. There is a long-standing tension in the muon AMM $(g-2)_\mu$ measured in 2006 at Brookhaven \cite{Bennett:2006fi}. Fermilab's new measurement \cite{Abi:2021gix} is in complete agreement with the previous result, and the combined result shows a large $+4.2\sigma$ discrepancy with
the SM prediction \cite{Aoyama:2020ynm}. For a recent review of BSM models addressing this anomaly, see Ref. \cite{Athron:2021iuf}. As for the electron, a recent experiment at Berkeley \cite{Parker:2018vye} has measured the fine-structure constant using Caesium atom with extreme precision, which indicates a $-2.4\sigma$ disagreement with the direct experimental measurement~\cite{Hanneke:2008tm}.  A large positive deviation  of $(g-2)_\mu$ but a negative deviation of $(g-2)_e$  has caused excitement in the high-energy physics community, and various new physics proposals have been made to solve these discrepancies simultaneously \cite{Giudice:2012ms, Davoudiasl:2018fbb,Crivellin:2018qmi,Liu:2018xkx,Dutta:2018fge, Han:2018znu, Crivellin:2019mvj,Endo:2019bcj, Abdullah:2019ofw, Bauer:2019gfk,Badziak:2019gaf,Hiller:2019mou,CarcamoHernandez:2019ydc,Cornella:2019uxs,Endo:2020mev,CarcamoHernandez:2020pxw,Haba:2020gkr, Bigaran:2020jil, Jana:2020pxx,Calibbi:2020emz,Chen:2020jvl,Yang:2020bmh,Hati:2020fzp,Dutta:2020scq,Botella:2020xzf,Chen:2020tfr, Dorsner:2020aaz, Arbelaez:2020rbq, Jana:2020joi,Chua:2020dya,Chun:2020uzw,Li:2020dbg,DelleRose:2020oaa,Kowalska:2020zve,Hernandez:2021tii,Bodas:2021fsy,Cao:2021lmj,Mondal:2021vou,CarcamoHernandez:2021iat,Han:2021gfu,Escribano:2021css,CarcamoHernandez:2021qhf,Chang:2021axw,Chowdhury:2021tnm,Bharadwaj:2021tgp,Borah:2021khc,Bigaran:2021kmn,PadmanabhanKovilakam:2022FN,Li:2021wzv,Biswas:2021dan,Barman:2021xeq,Chowdhury:2022jde}.

Here we point out that a more recent measurement utilizing Rubidium atom at the Kastler Brossel Laboratory in Paris \cite{Morel:2020dww} shows somewhat consistent  with the direct experimental measurement~\cite{Hanneke:2008tm}.  Contrary to the 2018 result by Berkeley National Laboratory \cite{Parker:2018vye}, this new 2020 result   \cite{Morel:2020dww} finds $\Delta a_e$ to be positive ($+1.6\sigma$), indicating a $\sim 5\sigma$  disagreement between these two experiments. 
Since these determinations by the two groups cannot be reconciled within the stated uncertainties, we, at first, will treat these two results as independent cases, presenting separate fits to demonstrate consistency of our model with each result. 
Later, however, we also perform a fit to illustrate how our model can be simultaneously  consistent  with the combined results of \cite{Parker:2018vye} and  \cite{Morel:2020dww}, favoring a Standard Model-like value.

Furthermore, LFU violating $B$-meson decays have been persistently observed in a series of experiments. Among these processes, the most prominent deviation has been observed in neutral-current transitions associated with the ratio: $R_{K^{(\ast)}}= Br\left(B\to K^{(\ast)}\mu^+\mu^-  \right)/Br\left(B\to K^{(\ast)}e^+e^-  \right)$,
which has very small theory uncertainties since hadronic uncertainties cancel out in the above ratios, making it extremely sensitive to new physics probes. In the SM, this ratio is predicted to be unity \cite{Descotes-Genon:2015uva,Bobeth:2007dw,Bordone:2016gaq,Straub:2018kue,Isidori:2020acz} due to the LFU property. However, experiments have  consistently found it to be somewhat smaller than one \cite{LHCb:2017avl,LHCb:2019hip,Belle:2019oag,BELLE:2019xld}; among them, the most precise measurement by LHCb \cite{LHCb:2021trn} alone quotes a significance of $3.1\sigma$ for $R_K$-ratio.\footnote{In addition to the clean observable $R_{K^{(\ast)}}$, several other observables are in tension with the SM predictions for example angular distributions and branching ratios of several  $b\to s\mu^+\mu^-$ modes \cite{LHCb:2015svh,LHCb:2015ycz,LHCb:2016ykl,Belle:2016fev,ATLAS:2018gqc,CMS:2015bcy,CMS:2017rzx,LHCb:2020lmf,LHCb:2020gog,LHCb:2015wdu}. When all these anomalies are combined, a global fit \cite{Capdevila:2017bsm,Aebischer:2019mlg} to data suggests $> 5\sigma$ discrepancy. In this Letter, we only focus on the  $R_{K^{(\ast)}}$ ratio; the rest of the observables suffer from large hadronic uncertainties, and we do not attempt to alleviate these discrepancies.} On the other hand, charged-current transitions in the $B$-meson decays  indicate an enhancement of the  ratio: $R_{D^{(\ast)}}= Br\left(B\to D^{(\ast)}\tau \overline \nu_\tau  \right)/Br\left(B\to D^{(\ast)}\ell \overline \nu  \right)$, with $\ell= e, \mu$. The ratios $R_{D^{(\ast)}}$ show deviations  \cite{BaBar:2013mob,Belle:2016dyj,LHCb:2015gmp,LHCb:2017rln,Belle:2019gij}  from the SM predictions \cite{Na:2015kha, Aoki:2016frl,Na:2015kha, Aoki:2016frl} with a combined significance of about $3\sigma$ \cite{HFLAV:2019otj}.

Besides all this, first and foremost, it is widely accepted that the SM cannot be the fundamental theory at all energies since it fails to incorporate neutrino oscillations firmly confirmed by various experiments; for reviews see Refs.~\cite{Cai:2017jrq,Cai:2017mow}. Motivated by this crucial drawback of the SM, in this Letter, we propose a simultaneous solution to all four of the anomalies mentioned above that are directly intertwined with the neutrino mass generation mechanism.  In this unified framework \cite{Julio:2022ton}, we employ two scalar leptoquarks\footnote{For a recent review on LQs, see Ref. \cite{Dorsner:2016wpm}. Moreover, 
for solutions to $B$-anomalies by employing leptoquarks (scalar and vector) see e.g. Refs. \cite{Tanaka:2012nw, Dorsner:2013tla, Sakaki:2013bfa, Duraisamy:2014sna,  Hiller:2014yaa, Buras:2014fpa,Gripaios:2014tna, Freytsis:2015qca, Pas:2015hca, Bauer:2015knc,Fajfer:2015ycq,  Deppisch:2016qqd, Li:2016vvp, Becirevic:2016yqi,Becirevic:2016oho, Sahoo:2016pet, Bhattacharya:2016mcc, Duraisamy:2016gsd, Barbieri:2016las,Crivellin:2017zlb, DAmico:2017mtc,Hiller:2017bzc, Becirevic:2017jtw, Cai:2017wry,Alok:2017sui, Sumensari:2017mud,Buttazzo:2017ixm,Crivellin:2017dsk, Guo:2017gxp,Aloni:2017ixa,Assad:2017iib, DiLuzio:2017vat,Calibbi:2017qbu,Chauhan:2017uil,Cline:2017aed,Sumensari:2017ovu, Biswas:2018jun,Muller:2018nwq,Blanke:2018sro, Schmaltz:2018nls,Azatov:2018knx, Sheng:2018vvm, Becirevic:2018afm, Hati:2018fzc, Azatov:2018kzb,Huang:2018nnq, Angelescu:2018tyl, DaRold:2018moy,Balaji:2018zna, Bansal:2018nwp,Mandal:2018kau,Iguro:2018vqb, Fornal:2018dqn, Kim:2018oih, deMedeirosVarzielas:2019lgb, Zhang:2019hth, Aydemir:2019ynb, deMedeirosVarzielas:2019okf, Cornella:2019hct,Datta:2019tuj, Popov:2019tyc, Bigaran:2019bqv, Hati:2019ufv, Coy:2019rfr, Balaji:2019kwe, Crivellin:2019dwb,  Cata:2019wbu,
Altmannshofer:2020axr, Cheung:2020sbq, Saad:2020ucl,Saad:2020ihm, Dev:2020qet,Crivellin:2020ukd,
Crivellin:2020tsz,Gherardi:2020qhc,
Babu:2020hun,Bordone:2020lnb,Crivellin:2020mjs,
Crivellin:2020oup,Hati:2020cyn,Angelescu:2021lln,Marzocca:2021azj,Crivellin:2021egp,Perez:2021ddi, Crivellin:2021ejk,Zhang:2021dgl,Bordone:2021usz,Carvunis:2021dss,Marzocca:2021miv,BhupalDev:2021ipu,Allwicher:2021rtd,Wang:2021uqz,Bandyopadhyay:2021pld,Qian:2021ihf,Fischer:2021sqw,Gherardi:2021pwm,Crivellin:2021lix, London:2021lfn,Bandyopadhyay:2021kue,Husek:2021isa,Afik:2021xmi,Heeck:2022znj,Julio:2022ton,Crivellin:2022mff,Chowdhury:2022dps}.} (LQs), $\{S_1(\overline 3,1,1/3), R_2(3,2,7/6)\}$,  which address the flavor anomalies and neutrino masses simultaneously. In this model, Majorana neutrino mass appears at two-loop order. Our novel proposal is minimal, and the \textit{only neutrino mass model} in the literature that resolves all of these aforementioned notable anomalies while satisfying low-energy flavor constraints and the LHC limits.\footnote{A few proposals of neutrino mass models attempted  to connect neutrino mass with flavor anomalies; however, these models address these anomalies only partially; for details, see Ref. \cite{Julio:2022ton}.} Furthermore, the proposed model is fully testable since a resolution of the charged-current $B$-anomaly requires the relevant LQ to have a mass around 1 TeV (the neutral-current and $g-2$ anomalies can be explained with relatively heavier LQ).

This paper is organized in the following way: we introduce the model in Sec.~\ref{sec:model} and discuss how anomalies are incorporated within this setup in Secs.~\ref{sec:observables} and \ref{sec:constraints}. After presenting our numerical results in Sec.~\ref{sec:result}, finally we conclude in Sec.~\ref{sec:conclusion}.

\section{Model}\label{sec:model}
In addition to the SM particle content, the proposed model  consists of two scalar LQs (SLQs), $S_1(\overline 3,1,+1/3)$, $R_2(3,2,7/6)$, and another BSM multiplet $\xi_3(3,3,2/3)$. We denote their component fields by
\begin{align}
R_2=\left( \begin{array}{c} R^{5/3} \\ R^{2/3} \end{array} \right),
\;S_1=S^{1/3},\;
\xi= \left( \begin{array}{cc} \frac{\xi^{2/3}}{\sqrt{2}} & \xi^{5/3} 
\\ \xi^{-1/3} & -\frac{\xi^{2/3}}{\sqrt{2}} \end{array} \right). \nonumber
\end{align}
The couplings of LQs with the SM fermions 
take the following form:
\begin{align}
\mathcal{L}_Y^{\rm new} &=  f^L_{ij} \overline u_{R i} R_2 \cdot L_{j} + f^R_{ij} \overline Q_{i} R_2\ell_{R j} 
\nonumber\\&
+ y^L_{ij} \overline{Q^c_i} \cdot L_{j} S_1 + y^R_{ij} \overline{u^c}_{R i} S_1 \ell_{R j},
\label{new-yuk}
\end{align}
where $Q$ and $L$ stand for, respectively, left-handed quark and lepton doublets, $u_R$ and $\ell_R$ are right-handed up-type quark and lepton, and  $i,j=1-3$ are family indices. We also use ``$\cdot$'' to denote $SU(2)$ contraction so that $L\cdot Q \equiv L^\rho Q^\sigma\epsilon_{\rho\sigma}$ with $\epsilon$ being the Levi-Civita tensor and $\rho,\sigma=$ being $SU(2)$ indices. Since explanations of flavor anomalies require that these LQs couple to quark-lepton bilinears, as usual, we turn off the diquark coupling of $S_1$ LQ, which guarantees baryon number conservation. Now, expanding these Yukawa terms, we obtain
\begin{align}
\mathcal{L}&=
\overline u_{R i}\left(f^L\right)_{ij}\ell_{L j} R^{5/3}+  \overline u_{R i}\left(-f^L\right)_{ij} \nu_{L j} R^{2/3}
\nonumber \\&
+\overline u_{L i} \left(f^R\right)_{ij} \ell_{R j} R^{5/3}
+\overline d_{L i} \left(V^\dagger f^R\right)_{ij} \ell_{R j}R^{2/3}
\nonumber \\&
+\overline{u^c}_{L i}  \left(y^L\right)_{ij}\ell_{L j} S^{1/3}
+\overline{d^c}_{L i} \left(-V^Ty^L\right)_{ij}\nu_{L j} S^{1/3}
\nonumber\\&
+\overline{u^c}_{R i} \left(y^R\right)_{ij} \ell_{R j} S^{1/3},\label{up-basis}
\end{align}
where, $V$ represents the CKM mixing matrix, and we have chosen to work in the so-called ``up-quark mass diagonal basis".

\begin{figure}[th!]
    \centering    \includegraphics[scale=0.4]{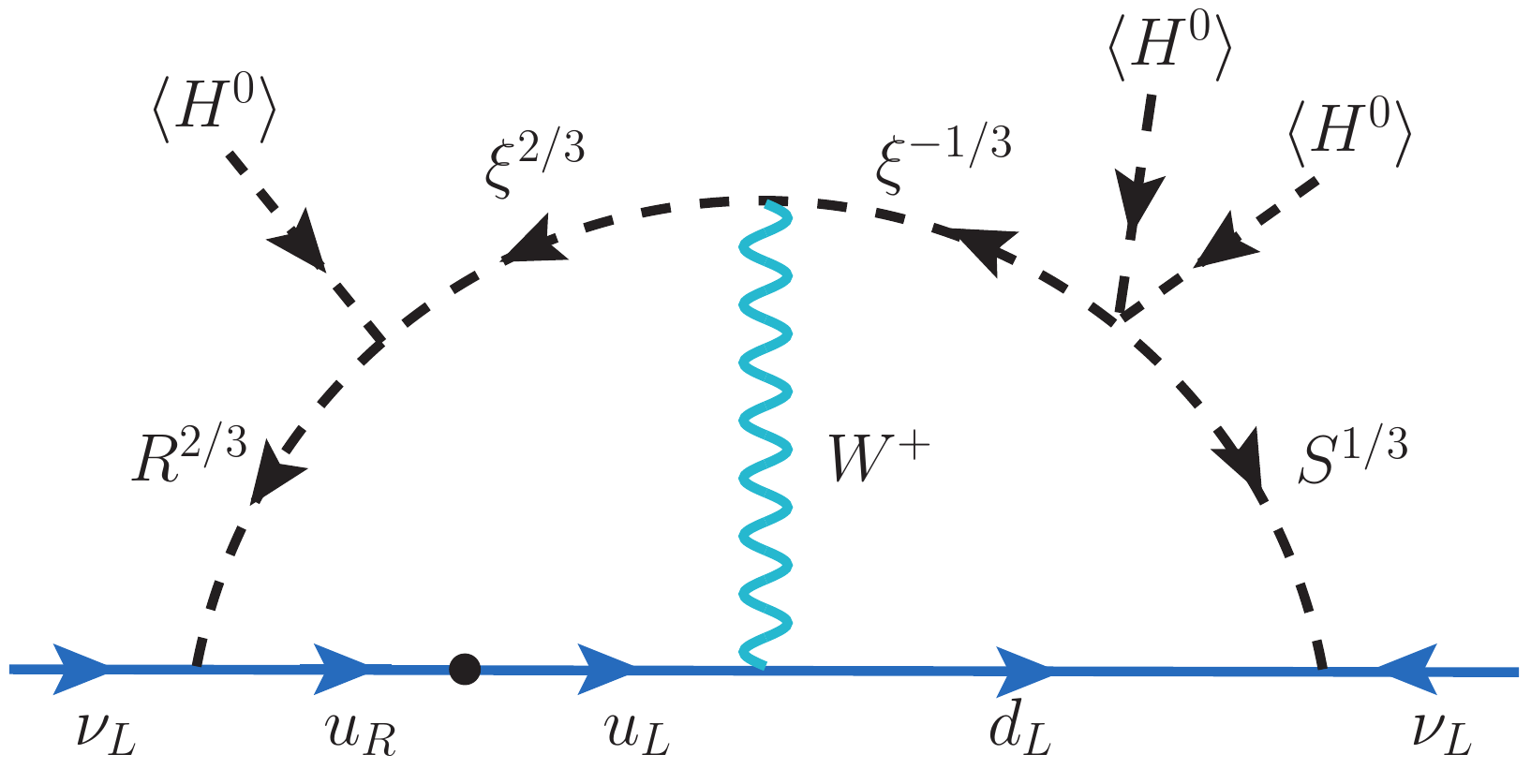}
    \caption{A typical two-loop diagram leading to non-zero neutrino mass. For full set of diagrams, see Ref.~\cite{Julio:2022ton}. }
    \label{Feynman}
\end{figure}

The two LQs and $\xi$ can also interact with Higgs doublet $H$ in the scalar potential $V_{sc} \supset  \lambda S_1^\dagger H^T\epsilon \xi^\dagger H + \mu R^\dagger_2 \xi H$.  The simultaneous presence of the Yukawa and these scalar interactions breaks the lepton number by two units, as required for generating non-zero neutrino mass; one of the diagrams is shown in Fig.~\ref{Feynman}. If the Yukawa interactions are turned off, the leptoquarks are not required to carry any lepton number; hence the scalar potential will remain lepton-number conserving.  After the EW symmetry breaking, these terms induce mixing in the LQ-$\xi$ sector, yielding six physical exotic scalars, i.e., $\chi^{1/3}_a$, $\chi^{2/3}_a$, and $\chi^{5/3}_a$ with $a=1,2$. Since explaining flavor anomalies requires some $\mathcal{O}(1)$ Yukawa couplings, we are compelled, by tiny neutrino masses, to have small mixing angles. Therefore, it is a good approximation to identify gauge eigenstates $R^{2/3,5/3}$ and $S^{1/3}$ as their physical eigenstates.

In the chosen basis, the neutrino mass formula takes the following form: 
\begin{align}
    {\mathcal M}_\nu = m_0I_0 
    \bigg\{& 2 (y^{L})^T D_u f^L + \frac{m_\tau}{m_t} D_\ell (y^{R})^Tf^L 
\nonumber \\
    &    
    - \frac{m_\tau}{m_t} D_\ell (f^{R})^T y^L \bigg\}   + {\rm transpose}, \label{nuEQ}
\end{align}
where $m_0=3g^2 m_t/\sqrt 2(16\pi^2)^2$ with $g$ being the $SU(2)$ gauge coupling constant and $m_t$ being the top-quark mass. The normalized mass matrices of up-type quarks and charged leptons are $D_u=\text{diag.}\left(\frac{m_u}{m_t}, \frac{m_c}{m_t}, 1 \right)$ and $D_\ell=\text{diag.}\left(\frac{m_e}{m_\tau}, \frac{m_\mu}{m_\tau}, 1 \right)$. The  loop integral $I_0$ in the asymptotic limit (i.e., when all fermion masses are zero)  is given by
\begin{align}
I_0 = \frac{1}{4}\sin2\theta\sin2\phi\sum_{a,b=1}^2(-1)^{a+b} I(M_{a+2},M_b). \label{eq4}
\end{align}
Here $\theta$ and $\phi$ denote the mixing angles of LQ with charges $1/3$ and $2/3$, respectively, while
\begin{align}
I(M_{a+2},M_b) = &~8\left(1-\frac{3}{4}\frac{M_{a+2}}{M_b}\right) - \frac{5}{6}\pi^2 \left( 1-\frac{2}{5}\frac{M_{a+2}}{M_b} \right) \nonumber \\
& + \frac{M_b^2}{m_W^2} \left(\frac{M_{a+2}}{M_b} -1 \right)\left(2-\tfrac{1}{3}\pi^2\right)  \nonumber \\
& + \left(\frac{2M_{a+2}}{M_b}-1 \right)\ln \frac{m_W^2}{M_b^2} \label{eq5}
\end{align}
is the loop function, derived by assuming $M_b/M_{a+2} =1+r_{ba}$ with $r_{ba}\ll 1$. Note that $M_{1,2}$ are masses for $2/3$ charged LQs and $M_{3,4}$ are for $1/3$ charged LQ. For more detailed discussion, we refer the reader to Ref.~\cite{Julio:2022ton}.

\section{Correlated Observables}\label{sec:observables}

\begin{table}[t!]
   { \footnotesize
    	{\renewcommand{\arraystretch}{1.5}%
    	\begin{center}
    \begin{tabular}{|c|c|}
    \hline
    \textbf{Process}&\textbf{Constraints}\\
    \hline
    \hline
        \multirow{2}*{$\mu\to e \gamma$ \cite{MEG:2016leq}}  &$|f_{\alpha e}^R f_{\alpha \mu}^{R* }|+|f_{\alpha e}^L f_{\alpha \mu}^{L* }|< 4.82\times 10^{-4}\left(\frac{M_{R_2}}{\text{TeV}}\right)^2$ \\
        \cline{2-2}
        & $(| f_{\alpha e}^R f_{\alpha \mu}^{L* } | + | f_{\alpha e}^L f_{\alpha \mu}^{R* } |)\ \mathcal{C} <7.63\times 10^{-5}$\ \\
        &\hspace{13mm}~~~~~~~$\hspace{5mm}~~~~~~~\left(\frac{M_{R_2}}{\text{TeV}} \right)^2 \left(\frac{1\ \text{GeV}}{m_q} \right) $ \\
        \hline
        \multirow{2}*{$\tau\to e \gamma$ \cite{BaBar:2009hkt}}  &$|f_{\alpha e}^R f_{\alpha \tau}^{R* }|+|f_{\alpha e}^L f_{\alpha \tau}^{L* }|< 0.32 \left(\frac{M_{R_2}}{\text{TeV}}\right)^2$ \\
        \cline{2-2}
        & $(| f_{\alpha e}^R f_{\alpha \tau}^{L* } | + | f_{\alpha e}^L f_{\alpha \tau}^{R* } |)\ \mathcal{C} <0.85 \left(\frac{M_{R_2}}{\text{TeV}} \right)^2 \left(\frac{1\ \text{GeV}}{m_q} \right)$\ \\
        \hline
        \multirow{2}*{$\tau\to \mu \gamma$ \cite{BaBar:2009hkt}}  &$|f_{\alpha \mu}^R f_{\alpha \tau}^{R* }|+|f_{\alpha \mu}^L f_{\alpha \tau}^{L* }|< 0.37 \left(\frac{M_{R_2}}{\text{TeV}}\right)^2$ \\
        \cline{2-2}
        & $(| f_{\alpha \mu}^R f_{\alpha \tau}^{L* } | + | f_{\alpha \mu}^L f_{\alpha \tau}^{R* } |)\ \mathcal{C} <0.98 \left(\frac{M_{R_2}}{\text{TeV}} \right)^2 \left(\frac{1\ \text{GeV}}{m_q} \right)$\ \\
        \hline
       $\mu - e$ \cite{SINDRUMII:2006dvw} &  $|\hat{f}_{ue}^R \hat{f}_{u\mu}^{R*}| \leq 8.58 \times 10^{-6} \left(\frac{M_{R_2}}{\text{TeV}}\right)^2$\\[3pt]
        \hline
    \end{tabular}
    \end{center}}
    }
    \caption{Constraints on Yukawa couplings from radiative decay of charged leptons. Here $\mathcal{C}=\Big(1+4 \log{\frac{m_q^2}{M_{R_2}^2}}\Big)$. The second (first) row in the ``Constraints" column shows the bound with (without) chiral enhancement. 
    Constraints on the $S_1$ LQ couplings $y^{L,R}$  with no chiral enhancement is weaker by a factor of 3, whereas chirally enhanced contribution has the factor with $C$ replaced by $C' = \Big(7+4 \log{\frac{m_q^2}{M_{R_2}^2}}\Big)$.  The contribution from $\hat{f}^R$ is suppressed as it is proportional to $m_b^2/M_{R_2}^2$ in comparison to $R_2^{5/3}$. }
    \label{tab:liljgamma}
\end{table}
This section discusses the relevant correlated observables/constraints in the proposed two-loop neutrino mass model that resolve the four anomalies.

At tree-level both  $S_1$ and $R_2$ LQs contribute to the $b\to c\tau\overline \nu$  transition which are favorable, whereas they do not lead to a desired contribution
to the $b\to s\mu\mu$. Loop-induced contributions to $b\to s\mu\mu$ transition face several difficulties and unable to consistently solve $R_{K^{(\ast)}}$ anomalies (see e.g. Ref.~\cite{Angelescu:2021lln}).  On the contrary, if $R_2$ LQ couples to the electrons, relatively small couplings  are feasible to address \cite{Popov:2019tyc,Altmannshofer:2021qrr} $R_{K^{(\ast)}}$ anomalies  since the relevant operator is generated at the tree-level. Moreover, each of these LQs can also simultaneously explain both the $(g-2)_\mu$ and $(g-2)_e$ anomalies via the chirally-enhanced top-quark and charm-quark contributions, respectively \cite{Dorsner:2020aaz}.

To be specific, $\hat f^R_{se,be}$ (for brevity, we define $\hat{f}^R \equiv V^\dagger f^R$ and $\hat{y}^L \equiv -V^T y^L$) couplings are responsible for addressing $R_{K^{(\ast)}}$ anomalies. Incorporating the $(g-2)_\mu$ ($(g-2)_e$) anomaly requires either $f^{R,L}_{t\mu}$ or $y^{R,L}_{t\mu}$ ($f^{R,L}_{t/ce}$ or $y^{R,L}_{t/ce}$) couplings to be sizable. Furthermore, with TeV scale LQ (LQ mass below TeV is highly constrained by LHC data), explanations of $R_{D^{(\ast)}}$ anomalies typically demand $\mathcal{O}(1)$  $y^R_{c\tau}$ and $\hat y^L_{b\ell}$ ($\hat f^R_{b\tau}$ and $f^L_{c\ell}$)  entries if $S_1$ ($R_2$) LQ is responsible for this charged-current processes.

Consequently, the NP contribution to resolving these four anomalies while incorporating neutrino oscillation data is highly non-trivial and leads the way to various flavor violating processes that are severely constrained by the experimental data. One of the most important constraints is the lepton flavor violating (LVF) radiative decay of charged leptons $\ell_i \to \ell_j \gamma$ \cite{Lavoura:2003xp}. For instance, the $\mu \to e \gamma$ process precludes $R_2$ LQ to accommodate $\Delta a_\mu$ while resolving $R_{K^{(*)}}$. Furthermore, to avoid strong constraints from chirally enhanced $\ell_i \to \ell_j \gamma$ via the CKM rotation in the up-sector, we have chosen to work in the up-quark mass diagonal basis of Eq.~\eqref{up-basis}. Constraints on the Yukawa couplings owing to dominant LFV processes as a function of LQ mass for both non-chiral and chirally enhanced diagrams are shown in Table~\ref{tab:liljgamma}  together with constraint from coherent $\mu - e$ conversion in nuclei.

Similarly, $R_2$ LQ in the up-quark diagonal basis leads to various  kaon decay processes~\cite{Mandal:2019gff,Ceccucci:2021gpl,Goudzovski:2022vbt} for which numerous stringent bounds are listed in Table~\ref{tab:kaon}. 
Moreover, the model requires $f^R_{b\tau}\sim \mathcal{O}(1)$, which modifies $Z$-boson decay to fermion pairs through one-loop radiative corrections. Then the LEP data \cite{ALEPH:2005ab} on the effective coupling leads to  $|f^R_{b\tau}| \leq 1.21$ \cite{Arnan:2019olv} for LQ mass of 1.0 TeV.

\vspace{3mm}
\textbf{Non-standard neutrino interactions:} 
Both $R_2$ and $S_1$ LQs have couplings to neutrinos and quarks that can induce charged-current NSI at tree-level~\cite{Wolfenstein:1977ue, Proceedings:2019qno, Babu:2019mfe}. 
Using the effective dimension-6 operators for NSI introduced in Ref.~\cite{Wolfenstein:1977ue}, the effective NSI parameters in our model are given by 
\begin{align}
    \varepsilon_{\alpha\beta} \ = \ \frac{3}{4\sqrt 2 G_F}
    \left(\frac{f^{L\star}_{u\alpha}f^L_{u\beta}}{M_{R^{2/3}}^2}
    +\frac{\hat{y}^L_{d\alpha}\hat{y}^L_{d\beta}}{M_{S^{1/3}}^2}
    \right) \, .
    \label{eq:NSI}
\end{align}
The Yukawa coupling $\hat y^L_{d\alpha}$ is subject to strong constraint from $K^+ \to \pi^+ \nu \nu$ as shown in Table~\ref{tab:kaon}. The couplings $f^L_{u\alpha}$ ($\alpha = e, \mu$) are constrained by the non-resonant dilepton searches at LHC and the limit on $f^L_{ue}$ and $f^L_{u\mu}$ are 0.19 and 0.16 for 1 TeV LQ mass. Thus, the contribution of $\hat y^L_{d\alpha}$ and $f^L_{u\alpha}$  which leads to $\varepsilon_{\alpha \beta}$  is at sub-percent level. However, the LHC limits on the LQ Yukawa coupling in the tau sector are weaker and in principle be $\mathcal{O} (1)$ leading to $\epsilon_{\tau \tau}$ as large as 34.4 \% \cite{Babu:2019mfe}, which is within reach of long-baseline neutrino experiments, such as DUNE \cite{Chatterjee:2021wac}.

\begin{table}[t]
{\footnotesize
    \centering
    \begin{tabular}{|c|c|}
    \hline
       {\bf Process}  & {\textbf{Constraints} \cite{Mandal:2019gff,Ceccucci:2021gpl,Goudzovski:2022vbt}}  \\[3pt]
       \hline\hline
       $K_L \to e^+ e^-$  & $|\hat{f}_{de}^R \hat{f}_{se}^{R*}| \leq 2.0 \times 10^{-3} \left(\frac{M_{R_2}}{\text{TeV}}\right)^2$ \\[3pt]
        \hline
        $K_L^0 \to e^\pm \mu^\mp$ & $|\hat{f}_{d\mu}^R \hat{f}_{se}^{R*}+ \hat{f}_{s\mu}^R \hat{f}_{de}^{R*}| \leq 1.9 \times 10^{-5} \left(\frac{M_{R_2}}{\text{TeV}}\right)^2$ \\[3pt]
        \hline
       $K_L^0 \to \pi^0 e^\pm \mu^\mp$ & $|\hat{f}_{d\mu}^R \hat{f}_{se}^{R*} - f_{s\mu}^R f_{de}^{R*}| \leq 2.9 \times 10^{-4} \left(\frac{M_{R_2}}{\text{TeV}}\right)^2$ \\[3pt]
        \hline
       $K^+ \to  \pi^+ e^+ e^-$  & $|\hat{f}_{de}^R \hat{f}_{s\mu}^{R*}| \leq 2.3 \times 10^{-2} \left(\frac{M_{R_2}}{\text{TeV}}\right)^2$\\[3pt]
        \hline
        $K^+ \to \pi^+ e^- \mu^+$ & $|\hat{f}_{d\mu}^R \hat{f}_{se}^{R*}|, |\hat{f}_{de}^R \hat{f}_{s\mu}^{R*}|    \leq 1.9 \times 10^{-4} \left(\frac{M_{R_2}}{\text{TeV}}\right)^2$ \\[3pt]
        \hline
        $K-\Bar{K}$ & $|\hat{f}_{d\alpha}^{R*} \hat{f}_{s\alpha}^{R}|  \leq 0.0266 \left(\frac{M_{R_2}}{\text{TeV}}\right)$ \\
        \hline
        $K^+ \to \pi^+ \nu \nu$ & Re[$\hat{y}^L_{de} \hat{y}^L_{se}$] $= [-3.7, 8.3] \times 10^{-4} \left(\frac{M_{S_1}}{\text{TeV}}\right)^2$ \\[3pt]
          & $[\sum_{m\neq n} |\hat{y}^L_{dm} \hat{y}^{L*}_{sn}|^2]^{1/2} < 6.0 \times 10^{-4} \left(\frac{M_{S_1}}{\text{TeV}}\right)^2$ \\[3pt]
        \hline
       $ B \to K^{(*)} \nu \nu$ & $\hat{y}^L_{b\alpha} \hat{y}^L_{s\beta} = [-0.036, 0.076]\ \left(\frac{M_{S_1}}{\text{TeV}}\right)^2$,\ [$R_{K^*}^{\nu \bar{\nu}} < 2.7$]\\[3pt]
        & $\hat{y}^L_{b\alpha} \hat{y}^L_{s\beta} = [-0.047, 0.087]\ \left(\frac{M_{S_1}}{\text{TeV}}\right)^2$,\ [$R_{K}^{\nu \bar{\nu}} < 3.9$]\\[3pt]
        \hline
    \end{tabular}
    }
    \caption{Constraints on the relevant LQ couplings as a function of mass from leptonic kaon and $B$ meson decays.  }
    \label{tab:kaon}
\end{table}

\begin{figure}[!t]
    \includegraphics[width=0.387\textwidth]{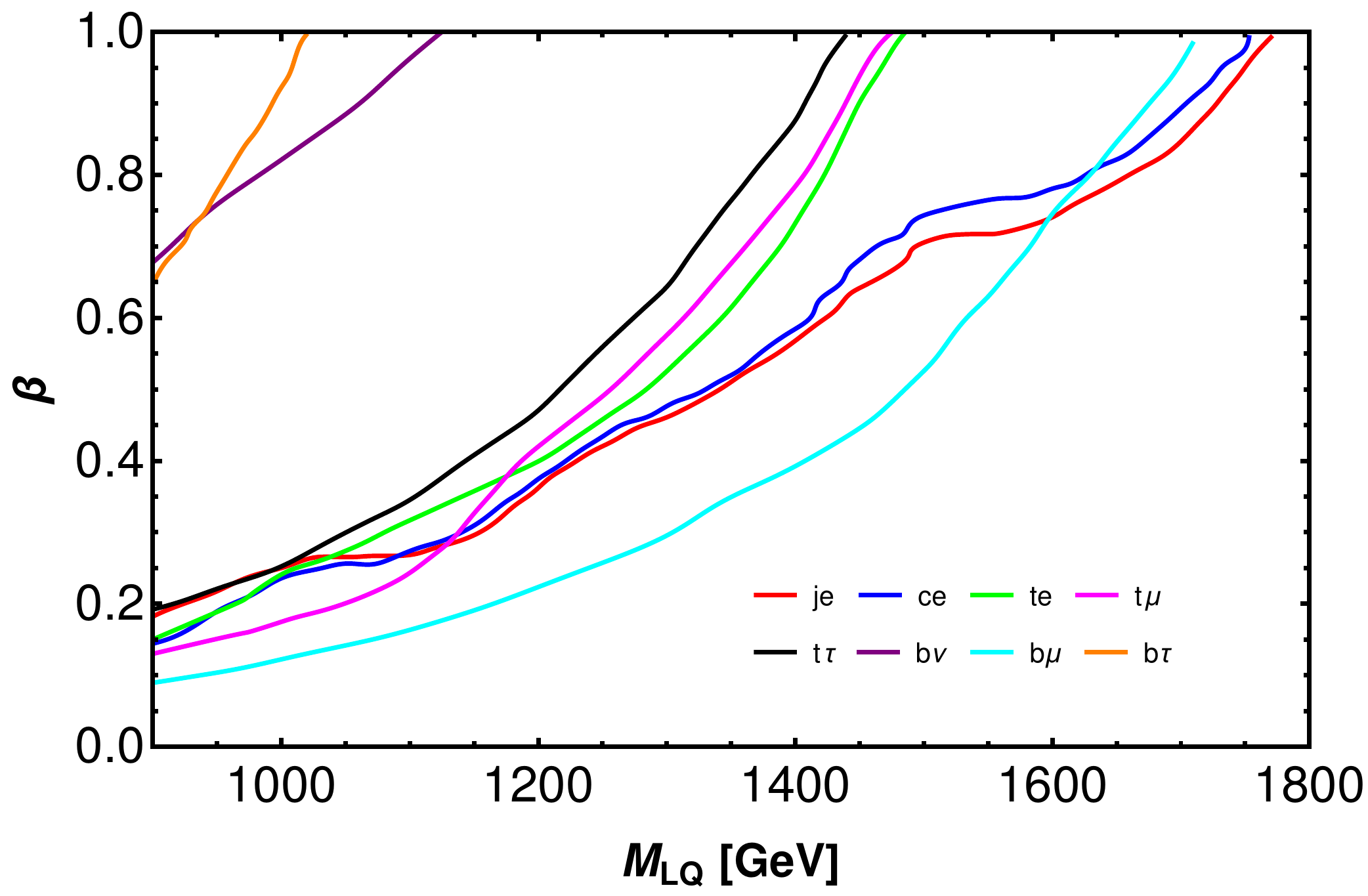}
    \includegraphics[width=0.377\textwidth]{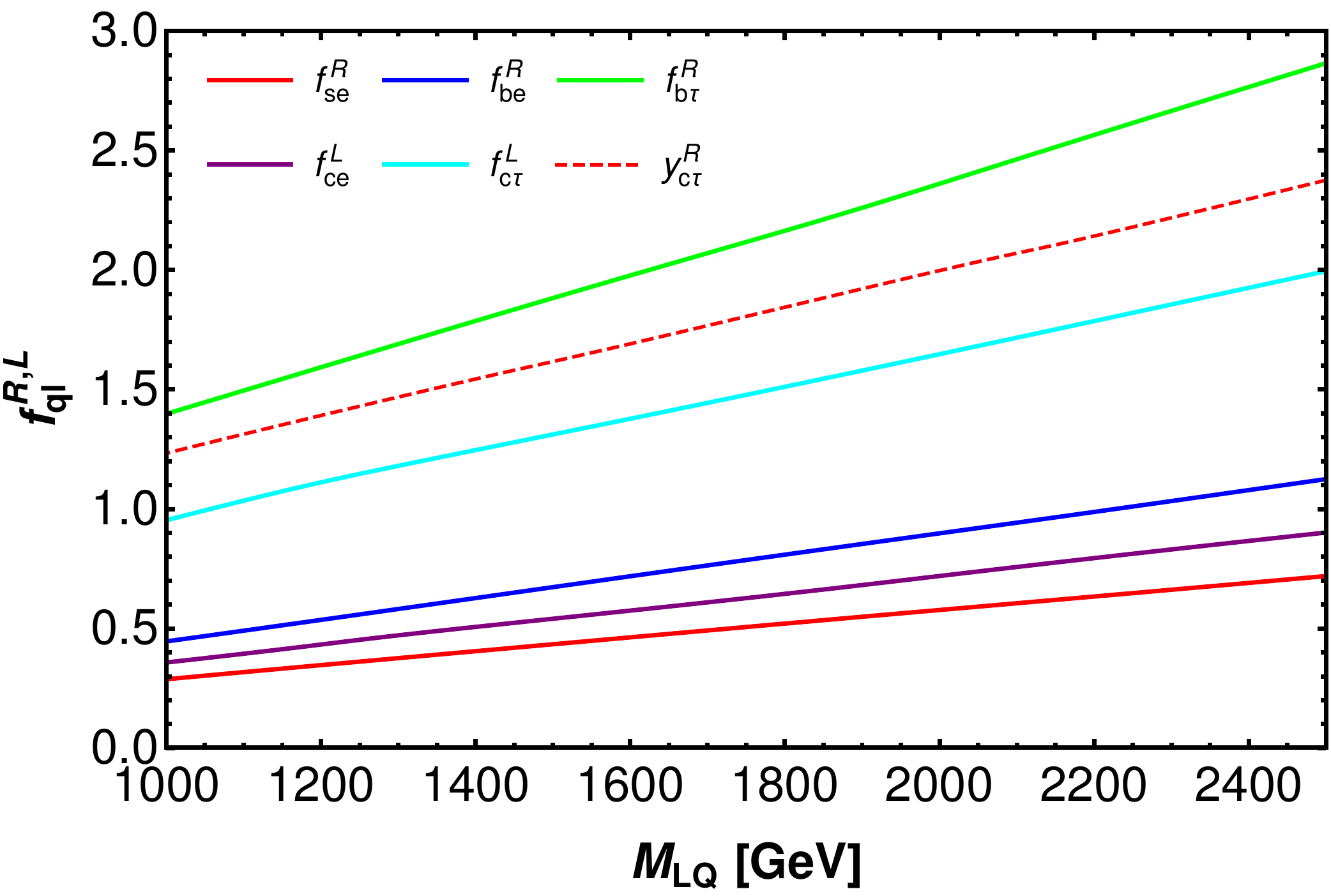}
    \caption{Upper plot: Summary of the current limits from searches for LQ pair-production  at the LHC for different quark-lepton final states. The branching ratio $\beta$ varies from 0 to 1 for a specific decay channel of the LQ, while the other decay channels that are not specified compensate for the missing branching ratios. Lower plot: Upper limits on the LQ couplings $f^{R,L}$ as a function of the LQ $R_2$ mass are presented, obtained from recent non-resonant dipleton searches at the LHC. The red dashed line represents bound on $y^R_{c\tau}$ for $S_1$ LQ.  }
    \label{fig:collider}
\end{figure}
\section{Constraints from Direct Searches}\label{sec:constraints}
At the LHC, $S_1$ and $R_2$ LQs can be pair produced \cite{Diaz:2017lit, Dorsner:2018ynv, Dorsner:2021chv} through $gg$ and $q\bar{q}$ fusion processes or can be singly produced in association with charged leptons via $s$- and $t$- channel quark-gluon fusion processes. The single production of LQ becomes only relevant for larger Yukawa couplings to the first and second-generation quarks \cite{Buonocore:2020erb}. Thus, collider bounds from single-production are not so significant compared to QCD-driven LQ pair production for our analysis. In both ATLAS and CMS, there are dedicated searches for the LQ pair production in different modes, LQ$^\dagger$ LQ $\to q\bar{q} \ell \bar{\ell}, q\bar{q} \nu \bar{\nu}$. The collider limits on LQ mass depend on the branching ratios to different modes, as depicted in Fig.~\ref{fig:collider} (upper-plot). Apart from the LQ-pair production bound, there are bounds on the couplings and mass on the LQ from the high-$p_T$ tails of $pp\to \ell \ell$ distributions \cite{Eboli:1987vb, Faroughy:2016osc, Schmaltz:2018nls, Greljo:2017vvb, Angelescu:2018tyl, Alves:2018krf, Angelescu:2020uug, Babu:2020hun, Angelescu:2021lln}. The corresponding bounds are also shown in Fig.~\ref{fig:collider} (lower-plot).

The above direct searches assume that the LQs decay promptly, which is no longer applicable for long-lived states. The colored particles arising from the $\xi$ field can generally be long-lived if their mixing with the LQs is sufficiently small, typically required to correctly incorporate the neutrino mass scale. Colored long-lived particles (CLLPs) hadronize prior to decay, forming
so-called R-hadrons~\cite{Farrar:1978xj} that are bound states composed of the CLLP and light quarks, antiquarks, and gluons. Since these particles carry color charge, for TeV scale masses, they are efficiently pair-produced $pp\to \chi_a\overline \chi_a$ via pure QCD interactions. Once produced, they decay to SM fermions with decay width $\Gamma\sim y^2 m_\chi/(16\pi)$~\cite{Buchmuller:1986zs} corresponding to a vertex displacement of $d\sim 10^{-14}/y^2$ mm. Therefore, for Yukawa couplings $y\gtrsim 10^{-6.5}$, decay is prompt since the corresponding vertex displacements are smaller than typical LHC detector resolution~\cite{Diaz:2017lit}. On the other hand, for $y\lesssim 10^{-6.5}$ they become long-lived, and for $y\lesssim 10^{-9}$ these states are practically stable and decay outside the detector. Long-lived gluinos and squarks have been extensively searched for at the LHC. However, no signal above the expected background is
found, which provides lower bounds on the CLLPs. Our scenario is similar to the long-lived sbottom ($Q_{em}=1/3$) and stop ($Q_{em}=2/3$) for which the current bounds at $95\%$ confidence level  are $m\geq 1250$ GeV~\cite{ATLAS:2019gqq} and  $m\geq 1800$ GeV~\cite{CMS:2020iwv}, respectively.

\section{Results and Discussion}\label{sec:result}
\begin{table}[!t]
{\footnotesize
\centering
\begin{tabular}{|c|c|c|c|}
\hline \hline
 {\textbf{Oscillation}}   &   {\textbf{$3\sigma$ range}} &  \multicolumn{2}{c|}{\textbf{Fit values}}\\
 \cline{3-4}
 {\bf parameters} & \textbf{{\tt NuFit5.1}}~\cite{Esteban:2020cvm} & TX I (NH) &  TX II (IH) \\ \hline \hline
 $\Delta m_{21}^2 (10^{-5}~{\rm eV}^2$)  &   6.82 - 8.04   &  7.41 & 7.39   \\ \hline
 $\Delta m_{23}^2 (10^{-3}~{\rm eV}^2) $(IH) &   2.410 - 2.574  & - &  2.53  \\ 
 $\Delta m_{31}^2 (10^{-3}~{\rm eV}^2) $(NH) &   2.43 - 2.593   & 2.53 & - \\ \hline
  $\sin^2{\theta_{12}}$   &   0.269 - 0.343 &   0.316 &  0.2986 \\ \hline
 $\sin^2{\theta_{23}}$  (IH) &   0.410 - 0.613  &  -  &  0.534  \\
 $\sin^2{\theta_{23}}$ (NH)  &   0.408 - 0.603  &  0.506  &  -  \\ \hline 
  $\sin^2{\theta_{13}} $  (IH) &   0.02055 - 0.02457  &  -  &  0.0227  \\
  $\sin^2{\theta_{13}}  $(NH)  &   0.02060 - 0.02435  &  0.0218  &-     \\\hline \hline
  \textbf{Observable} & \textbf{$1 \, \sigma$ range} &  &    \\ \hline \hline
$C_9^{ee} = C_{10}^{ee}$  & $[-1.65, -1.13]$ \cite{Altmannshofer:2021qrr} & $-1.39$ & $-1.57$  \\ \hline \hline
$(g-2)_e \, \, (10^{-14})$ & $-88 \pm 36 (48 \pm 30)$ & $-86 (48)$ & $-84 (60)$ \\ \hline
$(g-2)_\mu \, \, (10^{-10})$  & $25.1 \pm 6.0$  & 22.4 & 24.2 \\ \hline \hline
\end{tabular}
}
\caption{$3\sigma$ allowed ranges of the neutrino oscillation parameters are taken from a recent global-fit~\cite{Esteban:2020cvm}. $1\sigma$ allowed ranges of lepton $g-2$ and $C_{9}^{ee}=C_{10}^{ee}$ are also summarized. The last two columns contain the fit values of the observables for each texture. For $(g-2)_e$, the number in the parenthesis in the second  column corresponds to the measurement of \cite{Morel:2020dww}, and the corresponding fitted values are given in parentheses in the third and fourth columns for TX-I and TX-II, respectively. }
\label{tab:nu}
\end{table}
In this section, we present our numerical results and study the correlations among $R_{D^{(*)}}$, $R_{K^{(*)}}$, $\Delta a_\mu$, and $\Delta a_e$ anomalies within their $1\sigma$ measured values while consistently incorporating neutrino oscillations data. First, we demonstrate the viability of our proposed model by providing two benchmark points\footnote{In our numerical analysis, CKM matrix is expressed in terms of   Wolfenstein parametrization \cite{Wolfenstein:1983yz} with $\lambda = 0.2248$, $A=0.8235$, $\rho=0.1569$. For the low scale up-type quark masses,  we take $m_u (2 {\rm GeV}) = 2.16$ MeV, $m_c (m_c) = 1.27$ GeV, and $m_t (m_t) = 160$ GeV, which are then extrapolated to the LQ mass scale at 1 TeV \cite{Xing:2007fb}.} given in  Eqs.~\eqref{eq:TXI} and~\eqref{eq:TXII} for two distinct textures. 

\noindent{\bf TX-I:}
With $m_0 I_0 = 1.73\times 10^{-6}$ GeV, $M_{R_2} = 1.5$ TeV and $M_{S_1} = 1.2$ TeV: 
{\footnotesize
\begin{align}
    f^R &= \begin{pmatrix}
    0 & 0 & 0\\
    {\bf 0.13} & 0 & -0.027 \\
    {\bf 0.036} & 0.01 & 0
    \end{pmatrix} \, ,\hspace{5mm}
    f^L =  \begin{pmatrix}
    0 & 0 & {\it 1.0} \\
    0 & 0 & 6.36 \times 10^{-4} \\
    0 & 0 & -5.32 \times 10^{-6} 
    \end{pmatrix} \, , \nonumber\\[7pt]
    y^R &= \begin{pmatrix}
    0 & 0 & 0\\
     -0.3^* & 0 &  \fbox{1.0}  \\
    0 & 0.0025^\dagger & 0 
    \end{pmatrix} \, ,\hspace{5mm}
    y^L = \begin{pmatrix}
    0 & 0 & 0\\
   0.02^* & 0  & 0\\
    0 &  \fbox{1.2}^\dagger & 0
    \end{pmatrix} \, .
    \label{eq:TXI}
\end{align}
} 
This flavor structure explains $R_D - R_{D^*}$ ( \fbox{~} ), $(g-2)_\mu$ ( $~\dagger$ ), and $(g-2)_e$ ( $\star$ ) utilizing $S_1$ LQ, and $R_K - R_{K^*}$ ({\bf bold}) from $R_2$ LQ. Non-zero elements in black are the additional entries required to explain neutrino oscillation data. The only non-zero entry that induces observable NSI is shown in {\it italics}.\\

\noindent{\bf TX-II:} 
With $m_0 I_0 = 3.33 \times 10^{-6}$ GeV, $M_{R_2} = 1.0$ TeV and $M_{S_1} = 1.5$ TeV: 
{\footnotesize
\begin{align}
    f^R &= \begin{pmatrix}
    0 & 0 & 0\\
    {\bf -0.013^*} & 0 & 0 \\
    {\bf -0.180} & 0 & \fbox{-0.9\ i}
    \end{pmatrix} \, ,\hspace{3mm}
    f^L = \begin{pmatrix}
    0 & 0 & 0 \\
   0.3^* & 0 & \fbox{0.9} \\
    0 & 0 & 0  
    \end{pmatrix} \, , \nonumber\\[7pt]
    y^R &= \begin{pmatrix}
    0 & 0 & 0\\
    0 & 0 &  -1.85\times 10^{-4} \\
    0 & 1.0^\dagger & 0 
    \end{pmatrix} ,\hspace{1mm}
    y^L = 10^{-3} \begin{pmatrix}
    0 & 0 & 0\\
    -1.82 & 5.78\ i  & 0\\
    0 & 4.40^\dagger & 0
    \end{pmatrix} \, .
    \label{eq:TXII}
\end{align}
}
This flavor structure, on the other hand,  explains $R_D - R_{D^*}$ ( $\fbox{~}$ ), $R_K - R_{K^*} $ ({\bf bold}), and $(g-2)_e$ ( $\star$ ) by making use of  $R_2$  LQ, and $(g-2)_\mu$ ( $\dagger$ ) from $S_1$  LQ. For each of these textures, fitted neutrino observables  are presented in Table~\ref{tab:nu}, which is in excellent agreement with experimental data. The consistency of these benchmarks fits with the current  LHC bounds are discussed later in this section.

Fits presented in  Eqs.~\eqref{eq:TXI} and~\eqref{eq:TXII} incorporate the $(g-2)_e$ measurement of  Ref.~\cite{Parker:2018vye}. Here we clearly demonstrate how to consistently reproduce the more recent measurement of Ref.~\cite{Morel:2020dww}. Interestingly, for fit  Eq.~\eqref{eq:TXI}, the coupling $y^R_{21}$ does not play any role in neutrino mass generation. Hence reducing its size by a factor of $\sim 1/2$ and flipping its sign (to be precise, the new value is $y^R_{21}=0.17$)  reproduces the result of  Ref.~\cite{Morel:2020dww} without affecting any other observable. On the other hand, for fit Eq.~\eqref{eq:TXII} to replicate the measurement  of  Ref.~\cite{Morel:2020dww}, $f^R_{21}$ needs to be reduced while keeping its product with $f^R_{31}$ same. Moreover, the couplings $f^R_{21,31}$ play an  insignificant role in neutrino mass generation as it is always accompanied by electron mass (i.e., suppressed by a factor of $m_e/m_t$, see Eq.~\eqref{nuEQ}). For instance, $f^R_{21} = 0.0093$ and $f^R_{31} = 0.252$ correctly reproduce the fit values illustrated in Table \ref{tab:nu} as TX-II, while reproducing the $(g-2)_e$ result of Ref.~\cite{Morel:2020dww}.

It is worth noting that our model can, in fact, be simultaneously consistent with both values of $(g-2)_e$ inferred from Cs \cite{Parker:2018vye} and Rb \cite{Morel:2020dww} experiments. 
To illustrate this, we need to combine the two uncorrelated  values $a_{e,\rm Cs}\pm \sigma_{\rm Cs}$ and $a_{e,\rm Rb}\pm \sigma_{\rm Rb}$ of the same physical quantity $a_e$ to get the best estimate $a_e^{\rm comb.}\pm \sigma^{\rm comb.}$  via~\cite{Fabjan:2020wnt}
\begin{align}
&a_e^{\rm comb.}=\frac{\sum_{k} \frac{a_{e,k}}{\sigma^2_k}}{\sum_k \frac{1}{\sigma^2_k}}, \;\;\; \sigma^{\rm comb.}=    \frac{1}{\sqrt{\sum_k \frac{1}{\sigma^2_k}}},
\label{comb}
\end{align}
where $k=\rm Cs,~Rb$. For the reader's convenience, here, we collect those values of $a_e$ inferred from the measurements of the fine-structure constant: $a_{e,\rm Cs}= 0.00115965218161(23)$~\cite{Parker:2018vye}  and $a_{e,\rm Rb}=0.001159652180252(95)$~\cite{Morel:2020dww}. 
Using Eq.~\eqref{comb}, we obtain $a_e^{\rm comb.}=0.00115965218045(9)$. Comparing with the direct measurement,  i.e., $a_e^{\rm exp.}=0.00115965218073(28)$~\cite{Hanneke:2008tm}, we find a deviation 
\begin{align}
\Delta a_e^{\rm comb.}=a_e^{\rm exp.}-a_e^{\rm comb.}=(2.8\pm2.9)\times 10^{-13},    \label{comb-ae}
\end{align}
which clearly indicates that $a_e^{\rm exp.}$ is in a good agreement with the (combined) SM value. The error in Eq.~\eqref{comb-ae} is obtained by adding the errors of $a_e^{\rm comb.}$ and $a_e^{\rm exp.}$ in quadrature. Note that the combined value $\Delta a_e^{\rm comb.}$ leans toward $\Delta a_e^{\rm Rb}$ due to the much smaller uncertainty associated with the Rb experiment (hence $a_e^{\rm comb.}$ is heavily weighted by $a_e^{\rm Rb}$ in Eq.~\ref{comb}).   

Now, for TX-I with NH, one can turn off Yukawa coupling $y_R^{21}$ without changing the corresponding fit (see discussion above). This choice gives no new physics contribution to electron $(g-2)$, i.e., a special case corresponding to a pure SM-like scenario with $\Delta a_e=0$.  Furthermore, the combined value given in Eq.~\eqref{comb-ae} can also be trivially reproduced by setting $y_R^{21}=0.095$ without affecting any other observable. 

On the other hand, for IH with texture TX-II, setting $f^L_{21}=0$ would lead to the texture with $(\mathcal{M}_\nu)_{11} =(\mathcal{M}_\nu)_{22} = 0$, forbidden by the oscillation data.
One can however, lower the Yukawa coupling $f^R_{21}$ while keeping its product with $f^R_{31}$ fixed to obtain $C_9^{ee}=C_{10}^{ee}$ within $1\sigma$. Moreover, $f^R_{21} \gtrsim 0.0067$ is required to satisfy the bound $|f^R_{31} f^{R*}_{33}| < 0.32\ (M_{R_2}/{\rm TeV})^2$ obtained from $\tau \to e \gamma$. This choice of $f^R_{21}$ leads to $\Delta a_e \gtrsim 4.32 \times 10^{-13}$, consistent with combined measurement given in Eq.~\eqref{comb-ae}.

Next we show that a slight deviation from the TX-II opens up parameter space allowing even a smaller value of $\Delta a_e$. We demonstrate the viability by the following modified texture:  \\ \noindent{\bf TX-II$^{\prime}$:} 
With $m_0 I_0 = 4.77 \times 10^{-3}$ GeV, $M_{R_2} = 1.0$ TeV and $M_{S_1} = 1.5$ TeV: 
{\footnotesize
\begin{align}
    f^R &= \begin{pmatrix}
    0 & 0 & 0\\
    {\bf 0.026} & 0 & 0 \\
    {\bf 0.09} & 0 & \fbox{-0.9\ i}
    \end{pmatrix} \, ,\hspace{3mm}
    f^L = \begin{pmatrix}
    0 & 0 & 0 \\
    0.014 & 0 & \fbox{0.9} \\
    0 & 0 & 0.005\ i  
    \end{pmatrix} \, , \nonumber\\[7pt]
    y^R &= \begin{pmatrix}
    0 & 0 & 0\\
    0 & 0 &  0 \\
    0 & 1.0^\dagger & -0.132\ i 
    \end{pmatrix} ,\hspace{1mm}
    y^L = 10^{-3} \begin{pmatrix}
    0 & 0 & 0\\
    0.02 & 0.04  & -1.21\\
    0 & 4.40^\dagger & 0
    \end{pmatrix} \, .
    \label{eq:TXII}
\end{align}
}
The above benchmark, which is consistent with all flavor violating constrains, reproduces the central value of the neutrino observables {\tt NuFit5.1}~\cite{Esteban:2020cvm} and returns same values for the rest of the observables as in TX-II, except for $(g-2)_e$, which is found to be  $\Delta a_e = 7.7 \times 10^{-14}$. This number is also fully consistent with the combined $\Delta a_e$ given in Eq.~\eqref{comb-ae}.

\begin{figure}[t!]   
    \includegraphics[width=0.45\textwidth]{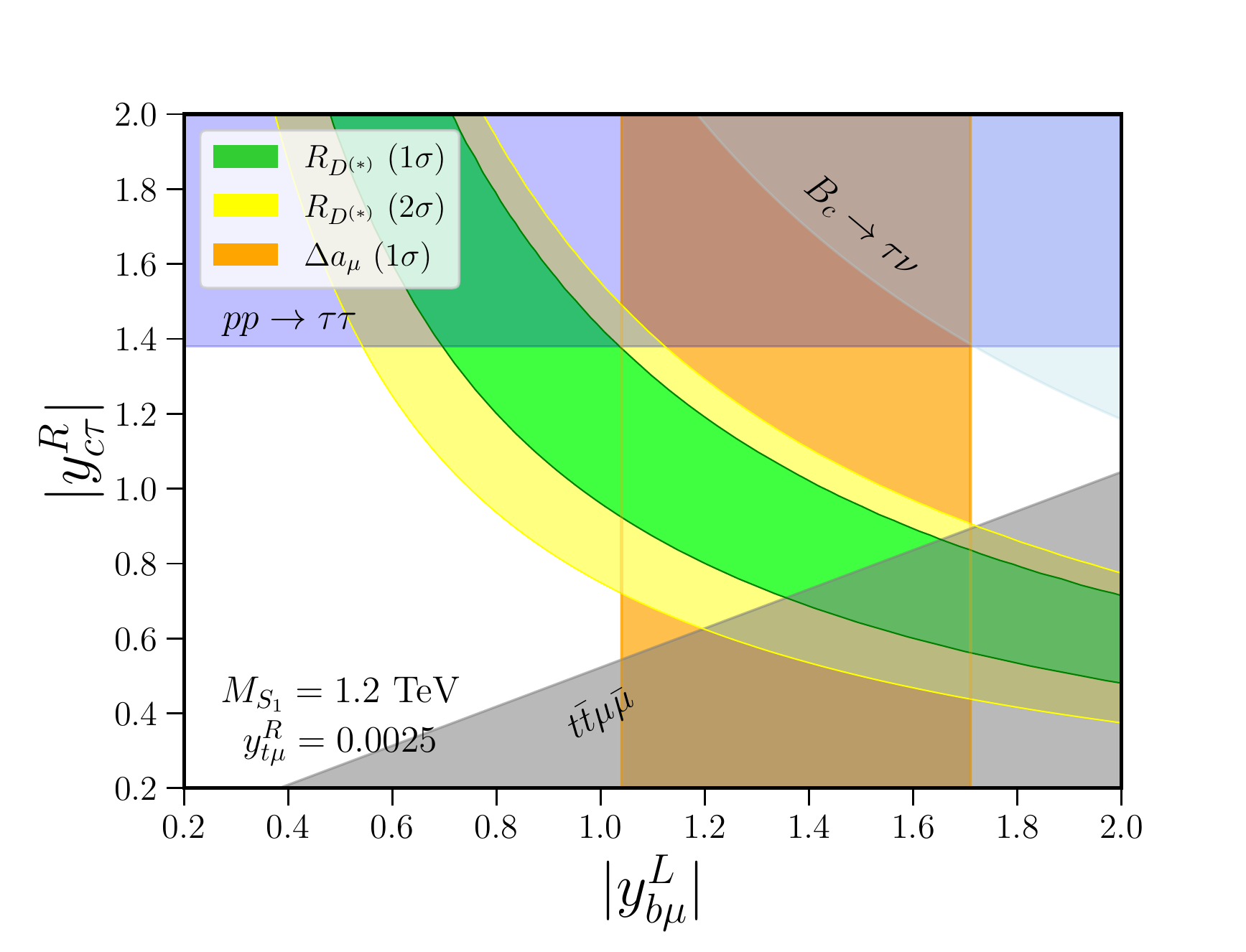}
     \includegraphics[width=0.45\textwidth]{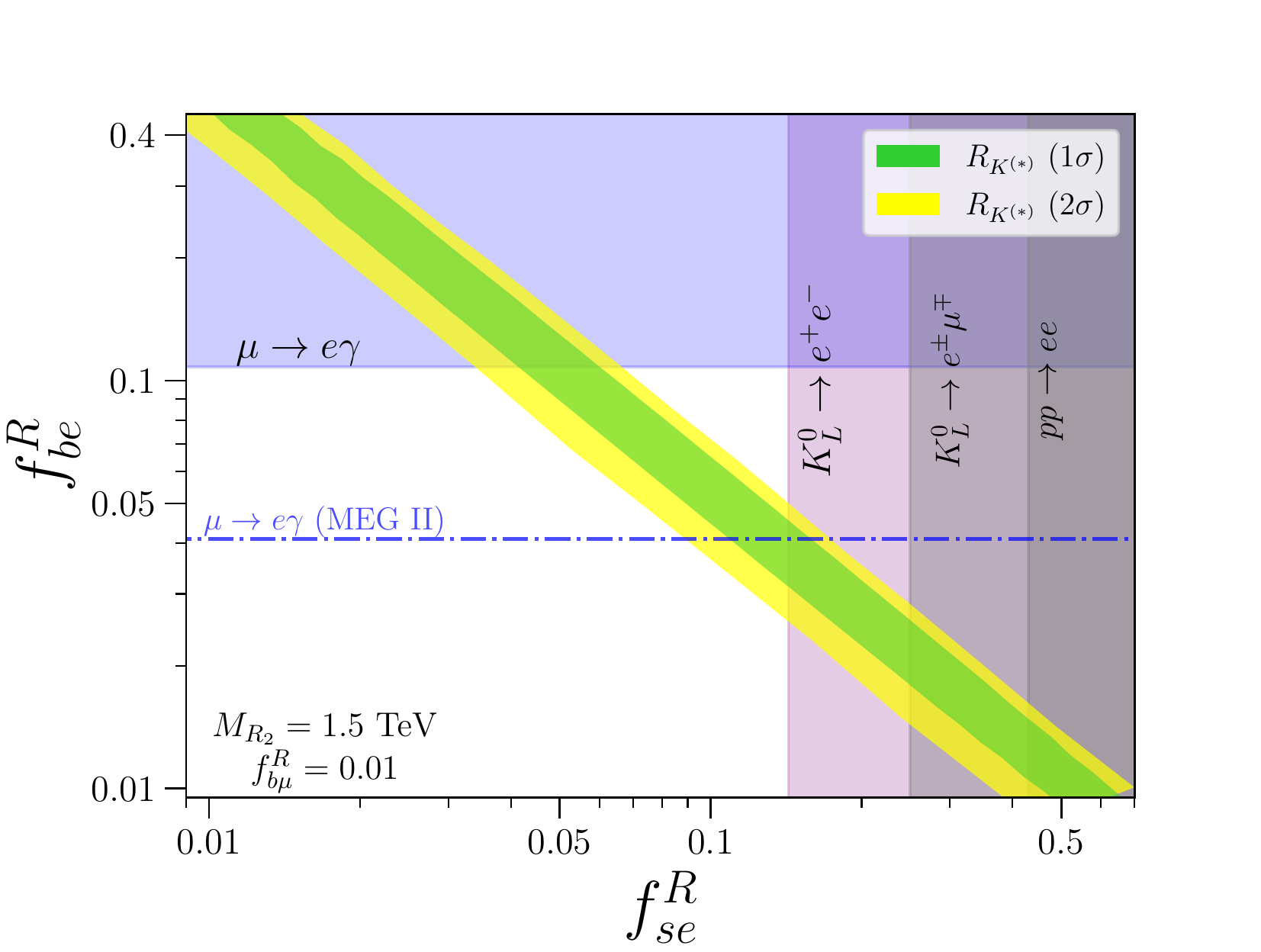}
    \caption{TX-I: $1\sigma$ (green) and $2\sigma$ (yellow) allowed range for $R_{D^{(*)}}$ (top) and $R_{K^(*)}$ (bottom) in the relevant Yukawa coupling planes, with the $S_1$ ($R_2$) LQ mass fixed at $1.2$ ($1.5$) TeV. Orange band on the top figure corresponds to $1\sigma$ allowed range of $\Delta a_\mu$. }
    \label{fig:TXI}
\end{figure}
Next, we study the correlations among numerous observables  associated with the benchmark scenarios, TX-I and TX-II.  For TX-I, the allowed parameter space to explain $R_{D^{(*)}}$ and $R_{K^{(*)}}$ at $1\sigma$ (green shaded) and $2\sigma$ (yellow shaded) CL in the relevant Yukawa coupling planes are illustrated in Fig.~\ref{fig:TXI} by fixing $R_2$ ($S_1$) mass at 1.2 (1.5) TeV. In computing $R_{D^{(*)}}$ and $R_{K^{(*)}}$ observables, we utilized {\tt Flavio package} \cite{Straub:2018kue}. By fixing $y^R_{t\mu} = 0.0025$ (cf. Eq.~\eqref{eq:TXI}), the allowed range to incorporate $\Delta a_\mu$ at $1\sigma$ is shown in orange band that corresponds to $y^R_{t\mu} y^L_{t\mu}\sim 10^{-3}$. We also show the exclusion regions obtained from resonant ($t\bar{t}\mu\bar{\mu}$) and non-resonant ($pp \to \tau \tau, e e$) production of LQs, along with flavor constraints such as $\mu \to e \gamma$ for a fixed $f^R_{b\mu}$, and kaon decays. Here the remaining anomaly $\Delta a_e$ is determined by the product $y^R_{ce} y^L_{ce}$, and the corresponding fitted value is listed in Table~\ref{tab:nu} obtained from the fit Eq.~\eqref{eq:TXI}.

\begin{figure}[t!]
    \includegraphics[width=0.45\textwidth]{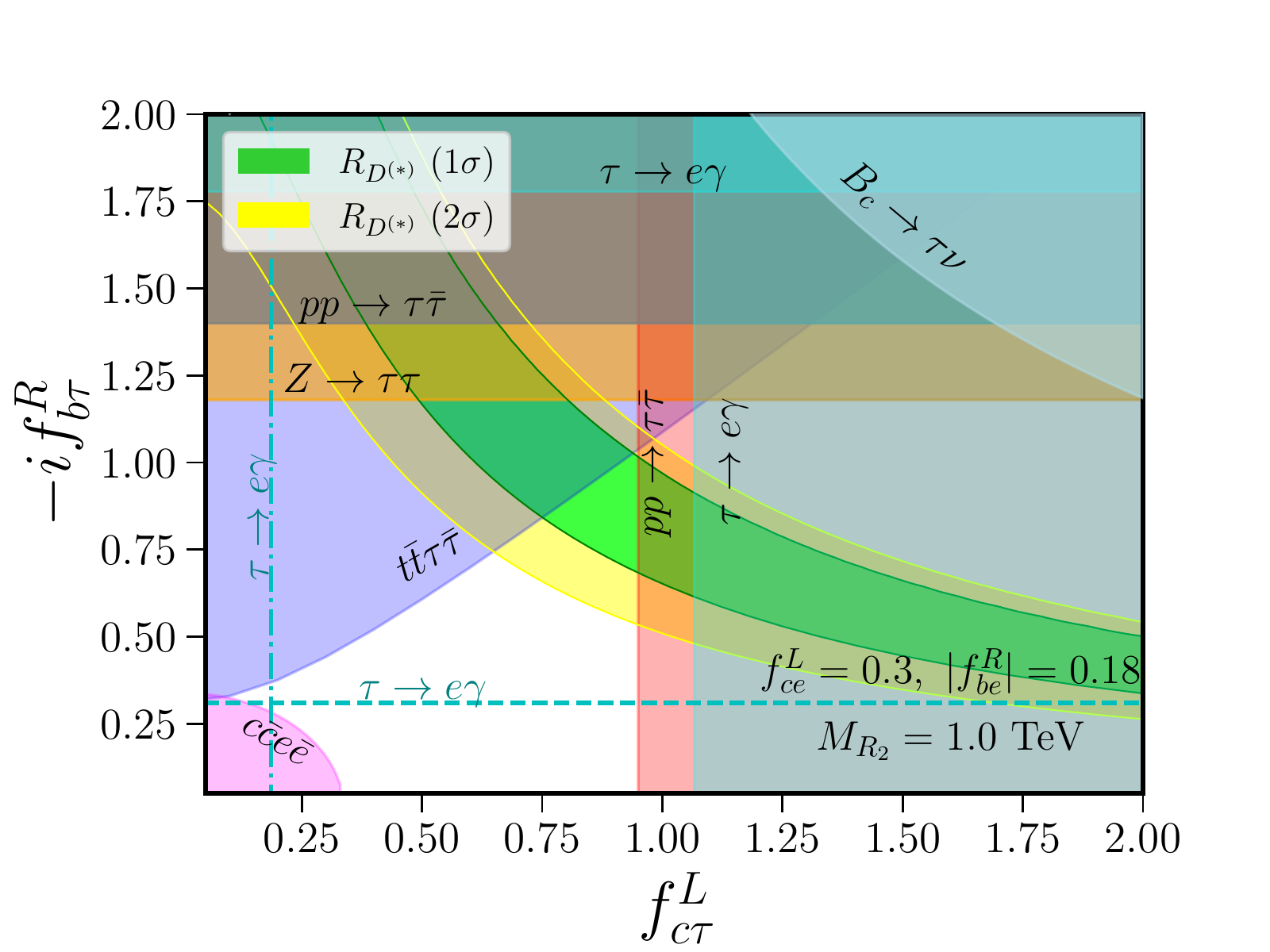}
     \includegraphics[width=0.45\textwidth]{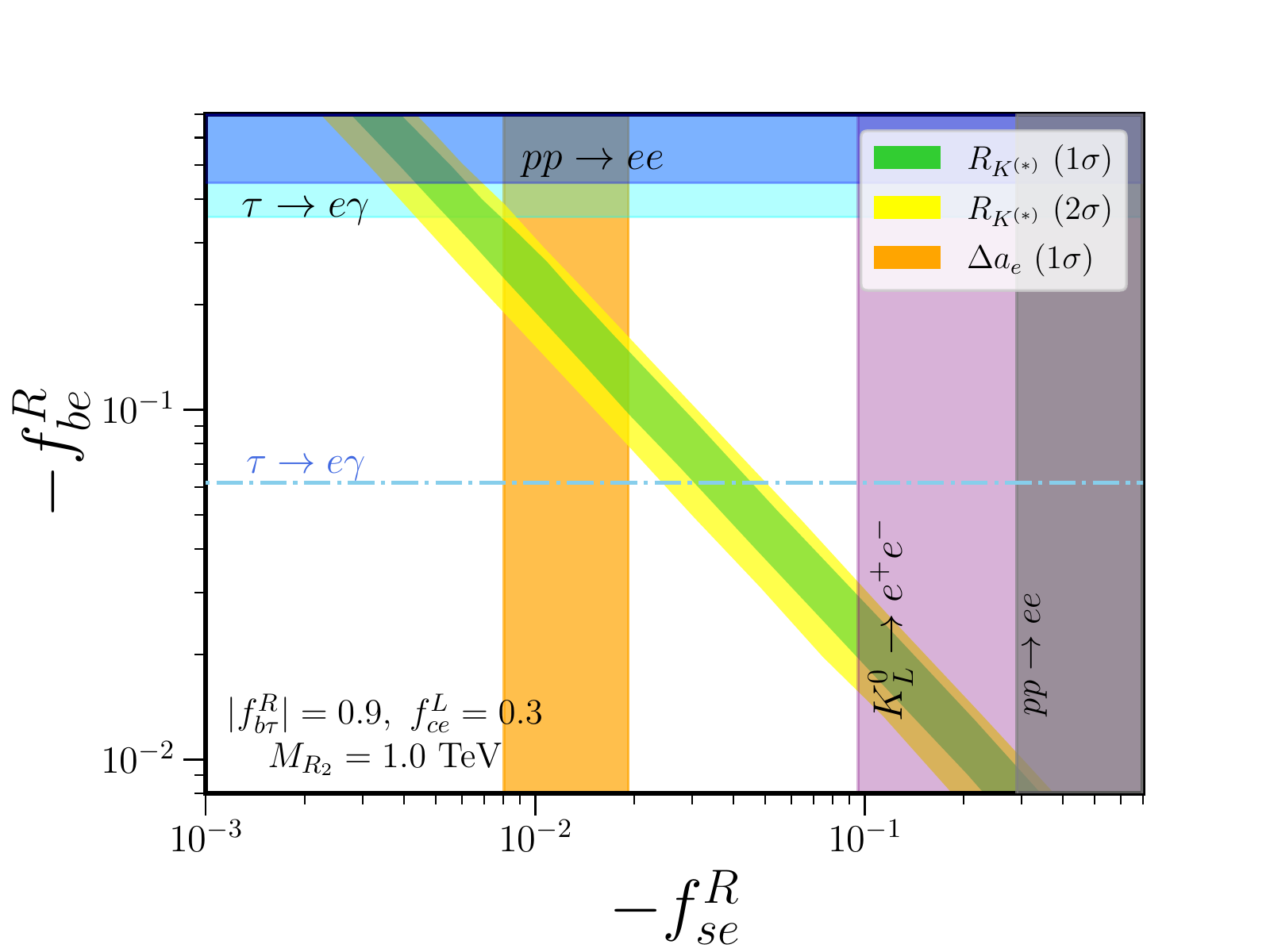}
    \caption{TX-II: $1\sigma$ (green) and $2\sigma$ (yellow) allowed range for $R_{D^{(*)}}$ (top) and $R_{K^(*)}$ (bottom) in the relevant Yukawa coupling planes, with the $R_2$ LQ mass fixed at $1.0$ TeV. Orange  band on the bottom figure corresponds to $1\sigma$ allowed range of $\Delta a_e$.  }
    \label{fig:TXII}
\end{figure}
Similarly, Fig.~\ref{fig:TXII} illustrates the TX-II scenario where $R_2$ LQ addresses $R_{D^{(*)}}$, $R_{K^{(*)}}$, and $\Delta a_e$ anomalies. Green and  yellow bands correspond to $1\sigma$ and $2\sigma$ allowed ranges for $R_{D^{(*)}}$ and $R_{K^{(*)}}$, respectively, whereas orange band shows   $1\sigma$ preferred value for $\Delta a_e$. In this plot, $M_{R_2} = 1$ TeV and $f^L_{ce} = 0.3$ (cf. Eq.~\eqref{eq:TXII}) are kept fixed.  The exclusion regions are obtained from resonant ($t\bar{t}\tau\bar{\tau}$, $c\bar{c}e\bar{e}$) and non-resonant ($pp \to \tau \tau, e e$) production of LQs, along with flavor constraints such as $\tau \to e \gamma$ for a fixed $f^R_{b\tau}$ and $f^R_{be}$, kaon decays, and $Z$-boson decay $Z\to \tau \tau$. $\Delta a_\mu$, which is not depicted in the figure, is determined by the product $y^R_{t\mu} y^L_{t\mu}$, and the corresponding benchmark value is shown in Table~\ref{tab:nu} obtained from  Eq.~\eqref{eq:TXII}.

As can be seen from these plots, the valid parameter space addressing all the anomalies is very limited due to various collider and flavor violating constraints, and upcoming experiments searching for LFV \cite{MEG:2016leq, Aushev:2010bq} have the potential to rule out this scenario. On the other hand, a resolution to flavor anomalies, particularly the $R_{D^{(\ast)}}$ ratios, requires LQ masses not higher than $\mathcal{O}(1)$ TeV, making this model fully testable at future high-luminosity LHC. 

\textbf{LHC bounds on fits TX-I and TX-II:} 
For TX-I with $f^L_{u\tau}=0$ (contributing to only NSI), $R_2$ LQ primarily decays to $jj e \bar{e}$  and $c \bar{c} e \bar{e}$ \cite{ATLAS:2020dsk} with the same Yukawa coupling $f^R_{se}$ leading to branching ratio  $\beta = 0.5$. Here we fix $R_2$ mass at $1.5$ TeV (corresponding to branching ratio $\beta \simeq 0.74$ to $jjee$). Moreover, once  $f^L_{u\tau}\sim \mathcal{O}(1)$  is included, new modes open up, and $R_2$ primarily decays to $jj\tau \bar{\tau}$ and $jj \nu \bar{\nu}$ \cite{CMS:2018qqq}. There are no dedicated searches for $jj \tau \bar{\tau}$ and the limits from $jj \nu \bar{\nu}$ is 1.0 TeV for $\beta =1$. On the other hand, for the TX-II, $R_2$ dominantly decays to $t\bar{t}\tau \bar{\tau}$ \cite{ATLAS:2021oiz}, $b \bar{b} \tau \bar{\tau}$ \cite{ATLAS:2019qpq, ATLAS:2021jyv}, $jj\tau\bar{\tau}$, and $j j \nu {\nu}$ each with the branching ratio $\beta \simeq 0.24$. Here we fix $R_2$ mass at 1 TeV, which allows the corresponding branching ratio to be as large as $\beta \simeq 0.27$ (from  $t\bar{t}\tau \bar{\tau}$).

On the other hand, for TX-I, $S_1$ primarily decays to $jj\tau\bar{\tau}$, $t \bar{t} \mu \bar{\mu}$ \cite{ATLAS:2020xov}, and $b \bar{b} \nu \bar{\nu}$ \cite{CMS:2018qqq} with the branching ratios (0.26, 0.37, 0.37), respectively. The latter two provide limits on the scalar LQ to be 1.5 (1.2) TeV and 1.1 (0.8) TeV for branching fraction $\beta = 1\ (\beta = 0.5)$, which is easily satisfied for our choice of $S_1$ LQ mass of 1.2 TeV. Similarly, for the TX-II $S_1$ decays to $t \bar{t} \mu \bar{\mu}$ $100\%$ of the time with the bound of 1.5 TeV as previously mentioned. For  TX-II, we fix $S_1$ LQ mass at 1.5 TeV  such that current collider constraints along with all the flavor constraints are easily satisfied. 

The bounds from dilepton final states ($q q \to \ell \ell$) \cite{ATLAS:2020zms, ATLAS:2020yat, CMS:2021ctt} on the relevant couplings $f^R_{se}, f^R_{be}, f^R_{b\tau}$, and $f^L_{c\tau}$ for $R_2$ LQ respectively read as $(0.29, 0.44, 1.4, 1.0)$ for 1 TeV mass and $(0.43, 0.67, 1.88, 1.31)$ for 1.5 TeV mass.  Similarly, for $S_1$ LQ the relevant Yukawa coupling $y^R_{c\tau}$ is constrained to be $1.39$ (1.61) for 1.2 (1.5) TeV mass.

Finally, we discuss the collider bounds on the states mostly originating from $\xi$. For our fits, using the definition of $m_0$ ($\sim 6\cdot 10^{-3}$ GeV), we find $I_0\sim 2\cdot 10^{-4}$. Moreover, for all physical scalars with masses close to each other (i.e., $r_{ba}\ll 1$), utilizing Eq.~\eqref{eq4}-\eqref{eq5}, we obtain $I_0\approx 2\sin 2\theta \sin 2 \phi$. Then, assuming mixings of similar order, we obtain $\theta\sim \phi\sim 10^{-2}$. Therefore, states arising primarily from $\xi$ have Yukawa couplings of order $10^{-2}$ with the third generation fermions (couplings with the first and second generations are more suppressed), hence decay promptly. Consequently, collider bounds on these states are identical to the bounds on LQs.

\section{Conclusions}\label{sec:conclusion}
In this Letter, we have presented for the first time a two-loop radiative neutrino mass model consisting of two scalars LQs $S_1(\overline 3,1,1/3)$ and $R_2(3,2,7/6)$ and another BSM scalar $\xi(3,3,2/3)$ with close-knit connections with $R_{K^{(\ast)}}$, $R_{D^{(\ast)}}$, and $(g-2)_{\mu, e}$ . We have performed a detailed analysis, including all relevant flavor violating and collider constraints, and presented benchmark fits showing complete consistency with anomalies and neutrino oscillation data. Furthermore, resolving these tantalizing flavor anomalies requires LQs to have masses around the TeV scale, enabling us to test this theory at the ongoing and future colliders. Finally, it is worth mentioning that the model sets several exciting avenues for future work; the implications on electric dipole moments and their correlations with the CP-violating phases in the neutrino sector, detailed analysis on LFV decays, and a dedicated collider study.

{\textbf {\textit {Acknowledgments.--}}} The work of J.J. was supported in part by the National Research and Innovation Agency of the Republic of Indonesia via Research Support Facility Program. 

\bibliographystyle{style}
\bibliography{references}
\end{document}